\documentclass[aps,twocolumn,superscriptaddress,showkeys,amsmath,amsfonts,
               floatfix]{revtex4}

\usepackage{epsfig,psfrag}
\usepackage{CJK}

\usepackage{graphicx}
\usepackage{dcolumn}
\usepackage{bm}
\usepackage{amssymb}

\begin{document}
\title{Correction of exciton binding energy in monolayer transition metal dichalcogenides}
\author{Zi-Wu Wang}
\affiliation{Tianjin Key Laboratory of Low Dimensional Materials Physics and Preparing Technology,
Department of Physics, Tianjin University, Tianjin 300354, China}
\email{wangziwu@tju.edu.cn}
\author{Yao Xiao}
\affiliation{Tianjin Key Laboratory of Low Dimensional Materials Physics and Preparing Technology,
Department of Physics, Tianjin University, Tianjin 300354, China}
\author{Wei-Ping Li}
\affiliation{Department of Basic Courses, Tianjin Sino-German University of Applied Sciences, Tianjin, 300350 People's Republic of China}
\author{Run-Ze Li}
\affiliation{Tianjin Key Laboratory of Low Dimensional Materials Physics and Preparing Technology,
Department of Physics, Tianjin University, Tianjin 300354, China}
\author{Zhi-Qing Li}
\affiliation{Tianjin Key Laboratory of Low Dimensional Materials Physics and Preparing Technology,
Department of Physics, Tianjin University, Tianjin 300354, China}
\begin{abstract}
\textbf{Abstract}: We theoretically investigate the corrections of exciton binding energy in monolayer transition metal dichalcogenides (TMDs) due to the exciton-optical phonon coupling in the Fr$\ddot{o}$hlich interaction model by using the linear operator combined Lee-Low-Pines variational method. We not only consider the excitons couple with the intrinsic longitudinal optical (LO) phonon modes, but also the surface optical phonon modes that induced by the polar substrates underneath the TMDs. We find that exciton binding energies are corrected in a large scale due to these exciton-optical phonon couplings. We discuss the dependences of exciton binding energy on the cut-off wave vector of optical phonon modes, the polarization parameters of materials and the interlayer distance between the polar substrates and TMDs. These results provide potential explanations for the divergence of the exciton binding energy between experiment and theory in TMDs.
\end{abstract}
\maketitle

\section{introduction}
Monolayer transition metal dichalcogenides MX$_2$ (TMDs) (Transition metal M=Mo, W; Chalcogen X=S, Se) have recently attracted great attention because of their distinctive electronic and optical properties differing from their corresponding bulk materials. These properties present a vast arena of potential applications\cite{wxj1,wx1} such as optoelectronics and valleytronics, quantum informatics and spintronics. One of the most significantly properties of these monolayer materials arising from the reduced dimensionality and dielectric screening is the strong Coulomb interaction between electrons and holes. This leads to the formation of tightly bound excitons and large exciton binding energies, which implies a rich excitonic physics in these monolayer materials\cite{wx1,wx2}.

There were a number of works have been focused on the excitonic properties of these monolayer materials in both experiment and theory, particularly on the exciton binding energies. In theoretical aspect, first-principle calculation based on GW and Bethe-Salpeter equation (GW-BSE) approach were used extensively to investigate the excitonic states and binding energy\cite{wx3,wx4,wx5,wx6,wx7}. The very large exciton binding energies in the range of 0.5$\sim$1 eV are predicted and confirmed by many consequent experiments such as the photoluminescence excitation spectroscopy\cite{wx8}, femtosecond transient absorption spectroscopy and microscopy\cite{wx9}, the optical reflection spectra\cite{wx10,wx11}, time-resolved photoluminescence spectroscopy\cite{wx12}. Within the frame of Wannier-Mott exciton, the simple hydrogenic model is also employed\cite{wx13,wx14}, in which the usual Rydberg series of energy levels of the excitonic states are obtained and applied to explain several experimental results successfully\cite{wx8,wx14,wx15}. However, the pronounced deviation from the usual hydrogenic model in the several lower excitonic levels were also observed in experiments\cite{wx16,wx17}. The modified Coulomb potential\cite{wx16,wx17} and effective two-dimensional dielectric constant\cite{wx18} models based on the nonlocal nature of dielectric screening are proposed to explain these deviations effectively.
\begin{figure}
\includegraphics[width=3.0in,keepaspectratio]{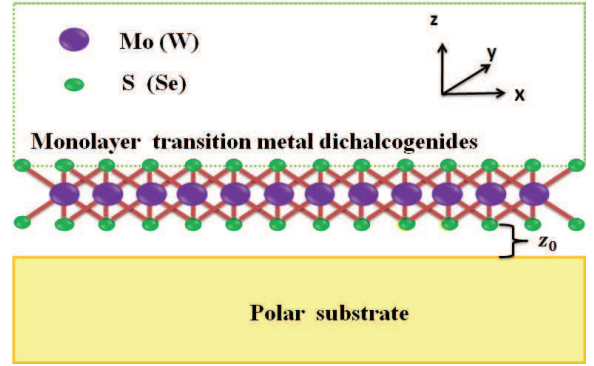}
\caption{\label{compare} Schematic diagram of the monolayer TMDs/polar substrates system. The monolayer material lies in the $x-y$ plane and $z_0$ is the distance between monolayer TMDs and polar substrate in the vertical direction $z$.}
\end{figure}
Recently, an analytically solvable model that the interband coupling between electron and hole states are considered to qualitatively explain nonhydrogenic exciton spectra observed in these monolayer TMDs\cite{wx19}. In addition, several exciton complexes and the corresponding binding energies are investigated by using the fractional dimensional space approach\cite{wx20}, in which the exciton binding energy depends on the modulation of the dimensionality parameter $\alpha$, which underlies the confinement effects of various geometrical structures. In fact that the  agreements between experiments and theories are obtained as above mentioned works. However, from the comparison of the obtained binding energies in experiments and theoretical models, it can be concluded that the causes regarding exciton binding energies can be varied in a large scale are still not clearly and the values of exciton binding energies for the same monolayer TMDs in various experiments and theoretical models can not get inconsistent. Therefore, the determining ingredients of exciton binding energies need to be explored further.

In the present paper, we mainly consider the influence of exciton-optical phonons couplings on the exciton binding energies in monolayer TMDs by using the linear operator and unitary transformation method. We not only consider the exciton couples with the intrinsic longitudinal optical (LO) phonon modes, but also the surface optical phonon modes that induced by the polar substrates under the TMDs. We find that the exciton binding energy can be varied from tens of meV to several hundreds of meV, depending on the cut-off wave vector of phonon modes, the polarization strength of substrates and the distance between the monolayer materials and polar substrates. These theoretical results highlight and enrich the understanding of the exciton states and exciton binding energy in these monolayer materials as well as the layered heterostructures.
\begin{figure}
\includegraphics[width=3.5in,keepaspectratio]{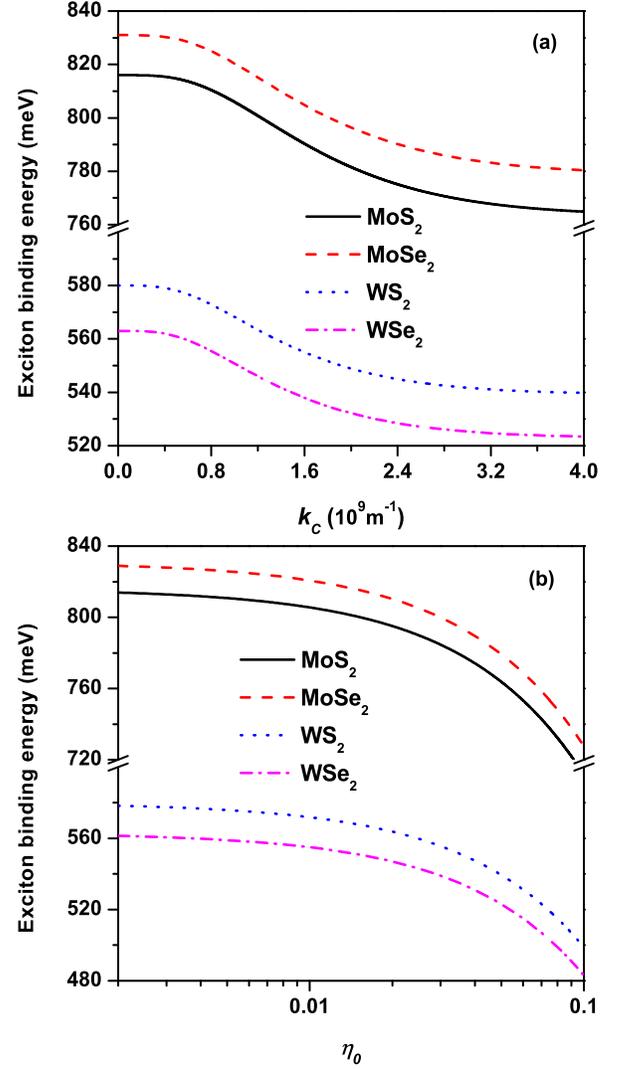}%
\caption{\label{compare} (a) Exciton binding energies of different TMDs as a function of the cut-off wave vector $k_c$ of LO phonon modes at $\eta_0$=0.05. (b) Exciton binding energies of different TMDs as a function of the polarization parameter $\eta_0$.}
\end{figure}

\section{theoretical model}
We consider the structure that the monolayer TMDs are placed on the polar substrate as schemed in Fig.1, in which the surface optical (SO) phonon modes are induced by the substrates. The exciton Hamiltonian of this structure taking into account the LO and SO phonons and the couplings between exciton and these optical phonons modes can be written as
\begin{subequations}
\begin{equation}
H=H_{ex}+H_{ph}+H_{ex-ph},
\end{equation}
\begin{equation}
H_{ex}=\frac{P^2}{2M}+\frac{p^2}{2\mu}-\frac{e^2}{4\pi\varepsilon_r r},
\end{equation}
\begin{equation}
H_{ph}=\sum_{k,\nu}\hbar\omega_{\nu} a_{k,\nu}^{\dagger} a_{k,\nu},
\end{equation}
\begin{eqnarray}
H_{ex-ph}&=&\sum_{k,\nu}[M_{k,\nu}\xi_k(r) a_{k,\nu}\exp{(ik\cdot R)}\nonumber\\
&&+M_{k,\nu}^*\xi_k^*(r) a_{k,\nu}^{\dagger}\exp{(-ik\cdot R)}].
\end{eqnarray}
\end{subequations}
The term H$_{ex}$ denotes the exciton Hamiltonian consists of an electron and a hole interacting with each other through a dynamically screened Coulomb potential ${e^2}/{(4\pi\varepsilon_r r)}$, $M=m_e+m_h$, $\mu=m_e^{-1}+m_h^{-1}$ are the center of mass, reduced mass, respectively. $m_e$ ($m_h$) is the electron (hole) effective mass, $r$ is the coordinate of relative motion, $P$ is the momentum of center of mass and $p$ is the momentum of relative motion. The second term $H_{ph}$ describes the optical phonon Hamiltonian, $\hbar\omega_{\nu}$ is the optical phonon energy, $\nu$=LO, SO$_1$, SO$_2$ stands for the LO mode and SO mode with two brabches, respectively. $a_k$ $(a_{k}^{\dagger})$ is annihilation (creation) a optical phonon with wave vector $k$. The third term $H_{e-ph}$ denotes the exciton-optical phonon coupling in the Fr$\ddot{o}$hlich interaction including two parts: one is the exciton-LO phonon coupling and the corresponding coupling element\cite{wx21,wx22,wx23}
\begin{equation}
M_{k,LO}=\sqrt{\frac{(e^2\eta_0 L_m\hbar\omega_{LO})}{(2A\varepsilon_0 k)}}erfc(\frac{k\sigma}{2}),
\end{equation}
in which $\eta_0$ is the polarization parameter and determined by the polarization properties of the monolayer TMDs, L$_m$ is the monolayer thickness, $erfc$ is the complementary error function and $\sigma$ denotes the effective width of the electronic Bloch states is based on the constrained interaction of LO phonon with charge carriers in monolayer materials, A is the quantization area in the monolayer plane, $\varepsilon_0$ is the permittivity of vacuum, R is the center of mass coordinate;
Another is the exciton-SO phonon coupling with the coupling element\cite{wx24,wx25,wx26}
\begin{equation}
M_{k,SO}=\sum_{\gamma=1,2}\sqrt{\frac{e^2\eta'\hbar\omega_{SO_{\gamma}}}{2A\varepsilon_0 k}}\exp{(-kz_0)},
\end{equation}
where the interlayer parameter $\eta'=(\kappa_0-\kappa_{\infty})/[(\kappa_{\infty}+1)(\kappa_0+1)]$ is a combination of the known dielectric constants of the substrates and denotes the polarization strength of polar substrate, $\kappa_{\infty}$ ($\kappa_0$) is the high (low) frequency dielectric constant. $z_0$ denotes the distance between the polar substrate and monolayer TMDs. The element $\xi_k(r)$ is expressed as
\begin{equation}
\xi_k(r)=\exp(-i\beta_1 k\cdot r)-\exp(-i\beta_2 k\cdot r),
\end{equation}
in which $\beta_1=m_e/M$ and $\beta_2=m_h/M$ is the fraction of electron and hole, respectively.

We introduce the creation $B_j^{\dagger}$ and annihilation $B_j$ operator for the center of mass momentum $P$ and coordinate $R$ by
\begin{subequations}
\begin{equation}
P_j=(\frac{M\hbar\lambda}{2}^{\frac{1}{2}}(B_j^{\dagger}+B_j),
\end{equation}
\begin{equation}
R_j=i(\frac{\hbar}{2M\lambda})^{\frac{1}{2}}(B_j-B_j^{\dagger}),
\end{equation}
\begin{equation}
(B_{j'}, B_j^{\dagger})=\delta_{j'j},
\end{equation}
\end{subequations}
and the well-known Lee-Low-Pines (LLP) unitary transformation by\cite{wx26,wx27,wx28}
\begin{subequations}
\begin{equation}
U_1=\exp\left(-i\sum_{k,\nu}a_{k,\nu}^{\dagger}a_{k,\nu}k\cdot R\right),
\end{equation}
\begin{equation}
U_2=\exp\left(\sum_{k,\nu}(F_{k,\nu} a_{k,\nu}^{\dagger} -F_{k,\nu}^{\ast}a_{k,\nu})\right),
\end{equation}
\end{subequations}
where the index $j$ refers to the $x$, $y$, $\lambda$ is the variational parameter and $F_k$ $(F_k^{\ast})$ is the variational function. $B_j^{\dagger}$ and $B_j$ are boson operators satisfying the boson commutative relation. Substituting Eq. (5a) and (5b) into the first term of Eq. (1b), and then performing the LLP transformation, we get
\begin{widetext}
\begin{eqnarray}
\widetilde{H}_{ex}&=&U_2^{-1}U_1^{-1}H_{ex}U_1U_2=H_{ex}^{(r)}+\frac{\hbar\lambda}{2}\sum_j(B_j^{\dagger}B_j+1)
+\frac{\hbar\lambda}{4}\sum_j(B_j^{\dagger}B_j^{\dagger}+B_jB_j)
+\hbar(\frac{\hbar\lambda}{2M})^{\frac{1}{2}}
\sum_{k,\nu}(a_{k,\nu}^{\dagger}+F_{k,\nu}^*)\nonumber\\
&&(a_{k,\nu}+F_{k,\nu})\sum_j k_j(B_j^{\dagger}+B_j)
+\frac{\hbar^2}{2M}\sum_{k,k',\nu}k\cdot k'(a_{k,\nu}^{\dagger}+F_{k,\nu}^*)(a_{k',\nu}^{\dagger}+F_{k',\nu}^*)
(a_{k,\nu}+F_{k,\nu})(a_{k',\nu}+F_{k',\nu})\nonumber\\
&-&\frac{\hbar^2}{2\mu}\sum_{k,k',\nu}(\nabla_r F_{k,\nu}\cdot \nabla_r F_{k',\nu}a_{k,\nu}^{\dagger}a_{k',\nu}^{\dagger}
+\nabla_r F_{k,\nu}^*\cdot \nabla_r F_{k',\nu}^*a_{k,\nu}a_{k',\nu})
+\frac{\hbar^2}{\mu}\sum_{k,k',\nu}\nabla_r F_{k,\nu}\cdot\nabla_r F_{k',\nu}^{*}a_{k,\nu}^{\dagger}a_{k',\nu}\nonumber\\
&+&\frac{\hbar^2}{2\mu}\sum_{k,\nu}|\nabla_r F_{k,\nu}|^2
-\frac{\hbar^2}{2\mu}\sum_{k,\nu}(a_{k,\nu}^{\dagger}\nabla_r F_{k,\nu}-a_{k,\nu}\nabla_r F_{k,\nu}^*)\cdot \nabla_r
-\frac{\hbar^2}{2\mu}\sum_{k,\nu}(a_{k,\nu}^{\dagger}\nabla_r^2 F_{k,\nu}-a_{k,\nu}\nabla_r^2 F_{k,\nu}^*),
\end{eqnarray}
\end{widetext}
with the Hamiltonian $H_{ex}^{(r)}=-\hbar^2\nabla_r^2/2\mu-e^2/4\pi\varepsilon_r r$ describes the relative motion of the electron-hole pair and be solved exactly in the hydrogenic model\cite{wx13,wx17,wx18}. The corresponding eigenfunction in poler coordinate (r,$\theta$) can be written as
\begin{eqnarray}
\phi_{n,l}&=&\frac{\zeta_n}{(2|l|)!}\left[\frac{(n+|l|-1)!}{(2n-1)(n-|l|-1)!}\right]^{\frac{1}{2}}
\frac{1}{\sqrt{2\pi}}\exp{(il\theta)}\nonumber\\
&&(\zeta_n r)^{|l|}\exp{(-\frac{\zeta_n r}{2})}\widetilde{F}(-n+|l|+1,2|l|-1,\zeta_n r),\nonumber\\
\end{eqnarray}
where $\zeta_n=\frac{2}{a_0(n-\frac{1}{2})}$, and $a_0=\frac{4\pi\varepsilon_r\hbar^2}{\mu e^2}$ is the Bohr radius, $\widetilde{F}(a,b,z)$ is the confluent hypergeometric function, and the quantum numbers $l$ takes the values $0,\pm1,\pm2,\ldots,\pm(n-1)$. The obtained binding energy of the $n$th excitonic state in this hydrogenic model is\cite{wx13,wx17,wx18}
\begin{equation}
E_b^{(n)}=\frac{1}{2(n-1/2)^2}\frac{e^2}{4\pi\varepsilon_r a_0}.
\end{equation}
This equation is extensively used to describe the exciton binding energy of $n$th excitonic state, in which the dielectric constant $\varepsilon_r$ plays a key role to determine the binding energy and related to many elements, e.g. the extension of excitonic radius, the reduced mass of exciton, the substrate environment. The nonhydrogenic Rydberg Series of excitonic states can also be well fitted via the modulation of parameter $\varepsilon_r$\cite{wx18}. In this paper, we adopted the values of $\varepsilon_r$ in Ref.30.

Substituting Eq. (5a) and (5b) into the term $H_{ph}$ and $H_{e-ph}$, and then performing the LLP unitary transformation, we have
\begin{widetext}
\begin{eqnarray}
\widetilde{H}_{ph}&=& U_2^{-1}U_1^{-1}H_{ph}U_1U_2
=\sum_{k,\nu}(\hbar\omega_{\nu}+\frac{\hbar^2 k^2}{2M})(a_{k,\nu}^{\dagger}a_{k,\nu} + a_{k,\nu}^{\dagger}F_{k,\nu} + F_{k,\nu}^{\ast}a_{k,\nu} + F_{k,\nu}^{\ast}F_{k,\nu}),
\end{eqnarray}
\begin{eqnarray}
\widetilde{H}_{ex-ph}&=& U_2^{-1}U_1^{-1}H_{e-ph}U_1U_2
=\sum_{k,\nu} \left[M_{k,\nu}^*\xi_k^*(r)(a_{k,\nu}^{\dagger}+F_{k,\nu}^{\ast})+M_{k,\nu}\xi_k(r)(a_{k,\nu}+F_{k,\nu})\right]
\exp{(-\frac{\hbar k^2}{4M\lambda})}\nonumber\\
&&\exp[-(\frac{\hbar}{2M\lambda})^{\frac{1}{2}}\sum_jk_jB_j^{\dagger}]
\exp[-(\frac{\hbar}{2M\lambda})^{\frac{1}{2}}\sum_jk_jB_j]
\end{eqnarray}
\end{widetext}
In order to obtain the binding energy of excitonic ground state, we choose $|\Phi\rangle=|\phi_{1,0}\rangle|0\rangle_{ex}|0\rangle_{ph}$  as the ground state wave-function of system. $|\phi_{1,0}\rangle$ is the ground state of exciton as given in Eq. (8), $|0\rangle_{ex}$ is the vacuum state vector of exciton describing the center of mass motion, $|0\rangle_{ph}$ is the zero phonon state. The exciton binding energy taking into account the exciton-optical phonon couplings can be obtained via
\begin{widetext}
\begin{eqnarray}
E(\lambda,F_{k,\nu},F_{k,\nu}^*)&=&\langle\Phi|\widetilde{H}_{ex}+\widetilde{H}_{ph}+\widetilde{H}_{ex-ph}|\Phi\rangle\nonumber\\
&=&E_b^{(0)}+\frac{\hbar\lambda}{2}+\sum_{k,\nu}(\hbar\omega_{\nu}+\frac{\hbar^2 k^2}{2M})F_{k,\nu}^*F_{k,\nu}
+\sum_{k,\nu}M_{k,\nu}^*F_{k,\nu}^*\exp{(-\frac{\hbar k^2}{4M\lambda})}\langle\phi_{1,0}|\xi_k^*(r)|\phi_{1,0}\rangle\nonumber\\
&+&\sum_{k,\nu}M_{k,\nu}F_{k,\nu}\exp{(-\frac{\hbar k^2}{4M\lambda})}\langle\phi_{1,0}|\xi_k(r)|\phi_{1,0}\rangle
+\frac{\hbar^2}{2\mu}\sum_{k,\nu}|\nabla_r F_{k,\nu}|^2,
\end{eqnarray}
\end{widetext}
in which the relations $B^{\dagger}_j|0\rangle_{ex}=|1\rangle_{ex}$, $B_j|0\rangle_{ex}=0$ and $a_k|0\rangle_{ph}=0$ are applied.

Minimizing Eq. (12) with respect to $f_k^{\ast}$ and $f_k$, we get
\begin{subequations}
\begin{equation}
f_k=-\frac{M_{k,\nu}\langle\phi_{1,0}|\xi_k^*(r)|\phi_{1,0}\rangle}{\hbar^2k^2/2M+\hbar\omega_{\nu}},
\end{equation}
\begin{equation}
f_k^{\ast}=-\frac{M_{k,\nu}\langle\phi_{1,0}|\xi_k(r)|\phi_{1,0}\rangle}{\hbar^2k^2/2M+\hbar\omega_{\nu}}
\end{equation}
\end{subequations}
Substituting Eq. (13a) and (13b) into Eq. (12), we get
\begin{eqnarray}
E_b&=&E_b^{(0)}+\frac{\hbar\lambda}{2}-
\sum_{k,\nu}\frac{|M_{k,\nu}|^2|\langle\phi_{1,0}|\xi_k(r)|\phi_{1,0}\rangle|^2}{\hbar^2k^2/2M+\hbar\omega_{\nu}}\nonumber\\
&-&\frac{\hbar^2}{2\mu}
\sum_{k,\nu}\frac{|M_{k,\nu}|^2|\langle\phi_{1,0}|\nabla_r\xi_k(r)|\phi_{1,0}\rangle|^2}{(\hbar^2k^2/2M+\hbar\omega_{\nu})^2},
\end{eqnarray}
in which each term is independent of $\lambda$ except the second term describing the center of mass movement. In order to clearly distinguish the influence of exciton-optical phonons coupling on the exciton binding energy, we set the case $\lambda=0$. Converting the summation of wave vector into the integral in Eq. (14), the exciton binding energy can be expressed as
\begin{eqnarray}
\widetilde{E_b}&=&E_b^{(0)}-
\sum_{\nu}\frac{A}{(2\pi)^2}\int_0^{k_c}[\frac{|M_{k,\nu}|^2\Re_1^2}{\hbar^2k^2/2M+\hbar\omega_{\nu}}\nonumber\\
&-&\frac{\hbar^2k^2}{2\mu}
\frac{|M_{k,\nu}|^2\Re_2^2}{(\hbar^2k^2/2M+\hbar\omega_{\nu})^2}]k dk\int_0^{2\pi}d\theta,
\end{eqnarray}
with
\begin{equation}
\Re_1=\frac{\zeta_0^3}{(\zeta_0^2+\beta_1^2k^2)^{3/2}}-\frac{\zeta_0^3}{(\zeta_0^2+\beta_2^2k^2)^{3/2}}\nonumber\\,
\end{equation}
\begin{equation}
\Re_2=\frac{\zeta_0^3}{(\zeta_0^2+\beta_1^2k^2)^{3/2}}+\frac{\zeta_0^3}{(\zeta_0^2+\beta_2^2k^2)^{3/2}}\nonumber\\,
\end{equation}
where $ k_c$ is the cut-off wave vector of optical phonon mode. The first term in integral denotes the motion of center of mass couples to the optical phonons. The correction of which to the exciton binding energy is very small, since the effective masses of electron and hole are very closer in these monolayer TMDs as listed in table I, leading to the polarization clouds of them cancel each other. The second term arises from the relative motion of exciton couples to the optical phonons, which gives the dominate contribution to the correction of the exciton binding energy and was assigned to the modification to the Coulomb potential in traditional exciton problems\cite{wx28}.
\begin{figure}
\includegraphics[width=3.5in,keepaspectratio]{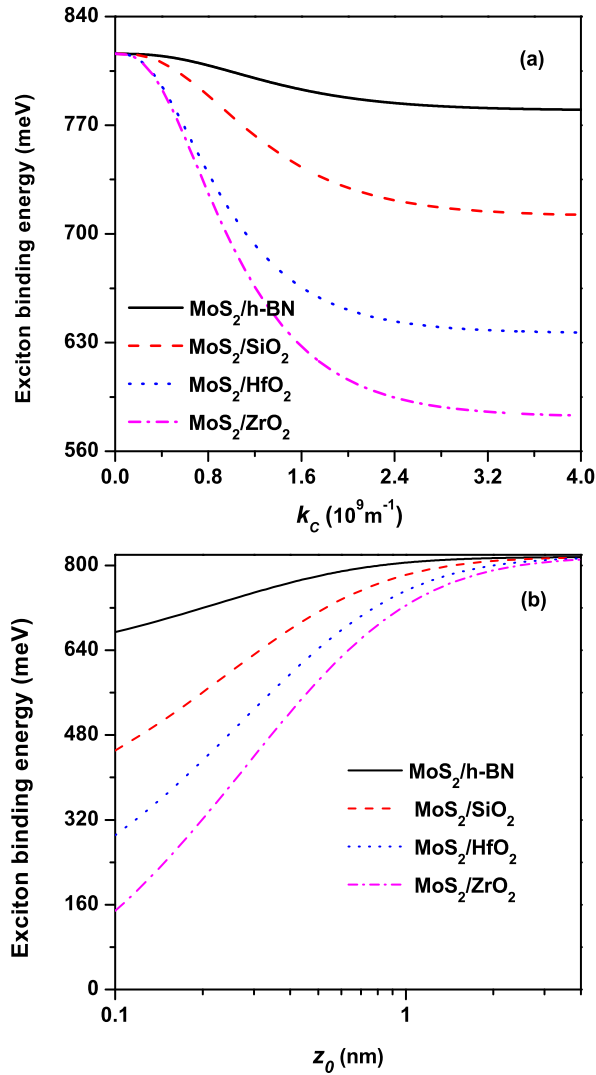}%
\caption{In the MoS$_2$/different polar substrates systems, (a) exciton binding energies as a function of the cut-off wave vector $k_c$ of SO phonon modes at $z_0$=0.5 nm; (b) exciton binding energies as a function of the distance $z_0$.}
\end{figure}
\begin{table}[htbp]
\caption{\label{compare}Parameters used to obtain results for Figs.2, 3 and 4. The effective masses of electron $m_e$ and hole $m_h$ (in units of the free electron mass $m_0$) are obtained from the Ref.31. The values of dielectric constant $\varepsilon_r$ and LO phonon energy are taken from Ref.30 and 32, respectively. Estimations for the exciton binding energies $E_b^{(0)}$ are obtained by using Eq. (12) in these monolayer TMDs.}
\begin{tabular}{cccccc}
\hline
                  &MoS$_2$ & MoSe$_2$ & WS$_2$ & WSe$_2$\\[1.0ex]\hline
$m_e$($m_0$)       &  0.51  &   0.64   &   0.31  & 0.39  \\
$m_h$($m_0$)       &  0.58  &   0.71   &   0.42  &  0.51  \\
$\varepsilon_r$($\varepsilon_0$)&  4.26  & 4.74 &  4.13  & 4.63 \\
$\hbar\omega_{LO}$(meV)  &  47  &   34   &   43  &  30   \\
$E_b^{(0)}$(meV)  &  816  &   831   &   580  &  563   \\
\hline
\end{tabular}
\end{table}
\begin{table}[htbp]
\caption{\label{compare}Surface-optical phonon modes and the dielectric constants of polar substrates used in this paper. Parameters have been taken from Refs.25 and 26.}
\begin{tabular}{ccccc}
\hline
&h-BN &  SiO$_2$  & HfO$_2$ &  ZrO$_2$\\[1.0ex]\hline
$\kappa_0$($\varepsilon_0$)       &  5.1  &   3.9   &   22.0  &  24.0  \\
$\kappa_{\infty}$($\varepsilon_0$)&  4.1  &   2.5   &   5.0  &  4.0  \\
$\hbar\omega_{SO,1}$(meV)         &  101  &   60   &   19  &   25   \\
$\hbar\omega_{SO,2}$(meV)         &  195  &   146   &  53  &  71   \\
\hline
\end{tabular}
\end{table}

\section{discussion and results}
The influence of the exciton-LO phonon coupling on the exciton binding energies are presented in Fig. 1 for different TMDs materials. We set the monolayer thickness of TMDs $L_m$=0.4 nm and the effective width of the electronic Block state $\sigma$=0.5 nm throughout the numerical calculation\cite{wx20,wx22}. Fig. 1(a) presents the dependence of exciton binding energies on the cut-off wave vector $k_c$ at $\eta_0=0.05$. As can be seen that exciton binding energies decrease with increasing the cut-off wave vector. When the $k_c$ reaches above the value of $4.0\times10^9m^{- 1}$, the effect of $k_c$ on exciton binding energies becomes insignificant, which means that only the phonons with small wave-vector, corresponding to the long-wave length phonons, are strongly coupled with the charge carriers, which is consistent with the results that small-momentum optical phonons give the mainly contributions to the Fr$\ddot{o}$hlich interaction based on first-principle calculation\cite{wx32}. Indeed, the influence of cut-off wave vector ($k_c$) of LO and SO phonon mode on the exciton binding energies are similar in following section, so we choose $k_c=4.0\times10^9m^{-1}$ as the limitation value for all numerical calculations presented in this paper. On the other hand, the strength of Fr$\ddot{o}$hlich interaction depends on the polarization properties of material, which is described by the polarization parameter $\eta_0$ in Eq. (2) and related to several ingredients, such as the dielectric tensor\cite{wx32}, screening of substrate material\cite{wx6}, thus leading to its precise estimation is a complex problem. We only qualitative analysis the influence of $\eta_0$ on exciton binding energies in Fig. 2(b) for different monolayer TMDs. One can see that the exciton binding energies can change a hundred of meV at $\eta_0=0.1$. Hence, the exciton binding energies are corrected remarkably if the larger values of $\eta_0$ obtained.

Inferred from Eqs. (3) and (15), it can be known that the corrections of exciton-SO phonon coupling to exciton binding energies depend on the cut-off wave vector $k_c$, polarization parameter of substrate $\eta'$ and distance $z_0$. These dependences are presented in Fig. 3. Surface optical phonon modes and dielectric constants of polar substrates are given in Table II, which have been taken from Refs. 25 and 26. In Fig. 3(a), exciton binding energies in monolayer MoS$_2$ for different polar substrate are presented at distance $z_0$=0.5 nm. Comparing with Fig. 2(a), it can be seen that the correction of exciton binding energies are enhanced more obviously with increasing $k_c$ due to two branches of SO phonon modes included. Moreover, the corrections increase from h-BN to ZrO$_2$ substrates, which because of the polarization parameter $\eta'$ that denotes the polarization strength of substrates increases from h-BN to ZrO$_2$ substrates as table II listed. The increasing of this parameter denotes the enhanced strength of exciton-SO phonons coupling, and thus resulting in the correction effect more stronger. Therefore, the polarization of substrate plays a significant role on the exciton binding energies in these monolayer materials. However, the coupling strength between exciton and SO phonons can be modulated by the distance $z_0$ as plotted in Fig. 3(b). From it, we can see that the corrections of exciton binding energies decrease sharply and can be varied in range of several hundreds of meV for different polar substrate with increasing the distance $z_0$. Moreover, the exciton binding energies change hardly when this distance exceeding $z_0$=3.0 nm, which means that the contribution of SO phonons to correction are very minor or negligible above this distance and LO phonons dominate the correction of exciton binding energy. This also provides the highlight for how to evaluate and control the influence of SO phonons induced by the substrates on the exciton states in experiments. In addition, the screening effect of substrate also leads to the large reduction in exciton binding energy has been calculated based on the first-principle\cite{wx6}. Although both this screening effect and SO phonons are related to the polar substrates and give potential correction to exciton binding energy, the correlations between them  can not discuss quantitative in the present study.
\begin{figure}
\includegraphics[width=3.5in,keepaspectratio]{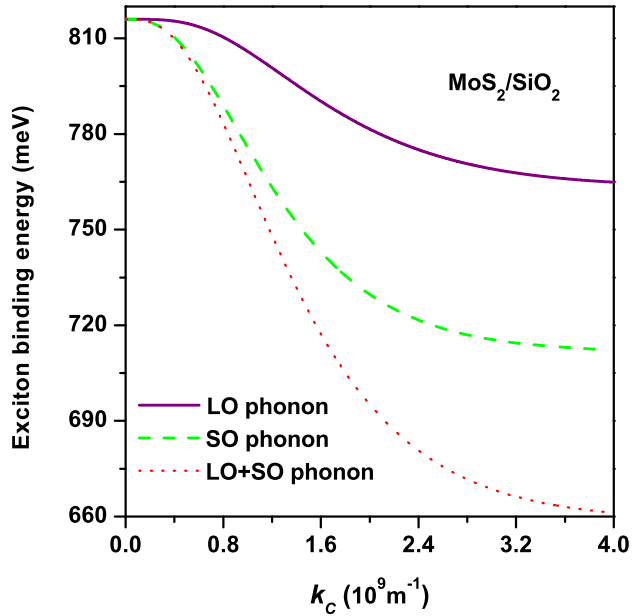}%
\caption{Exciton binding energy as functions of the cut-off wave number of optical phonon including LO and SO modes in MoS$_2$/SiO$_2$ system at the distance z$_0$=0.5 nm and the parameter $\eta_0=0.05$.}
\end{figure}

The total corrections of exciton-LO and SO phonons coupling on exciton binding energies for the monolayer MoS$_2$/SiO$_2$ system are presented in Fig. 4. It can be seen that the magnitude of exciton binding energy can be tuned in a very large scale due to this two kinds of exciton-optical phonons coupling, which not only provides an potential explanation for the divergence of the exciton binding energy between experiment and theory, but also proposes a way to identify the variation of exciton binding energy by controlling the polarization of substrate, the distance $z_0$ and dielectric environment in experiment. In fact, many-body mechanisms have been assigned to result in these discrepancies in several works\cite{wx5,wx6,wx13}. The present model proved that exciton-optical phonons coupling is an effective many-body interaction affecting exciton binding energies remarkably as plotted in Fig. 4. Especially, the SO phonons induced by the polar substrates have potential influence on the properties of excitonic states at certain distance $z_0$. These SO phonons modes generate a series of novel quantum phenomena in the van der Waals heterostructures based on these monolayer TMDs materials and arouse more and more interesting in recent years\cite{wx33}.

In summary, we theoretically investigate the corrections of exciton binding energies in monolayer TMDs by considering the exciton couples with LO phonons and SO phonons that induced by polar substrates. We find that (1) exciton binding energies are corrected in a large scale due to exciton-optical phonon coupling and those long-wave length phonons give the dominate contributions to the correction; (2) the polarization parameter of LO phonon and the polarization of substrates play a significant role in determining the magnitude of correction; (3) the influence of SO phonons become very weak when the distance between monolayer TMDs and substrates exceeds the range of $\sim$ 3 nm. These results highlight the discrepancies between theory and experiment regarding exciton binding energies in these monolayer TMDs.

This work was supported by National Natural Science Foundation of China (NO. 11304355 and NO. 11674241) and Young Core Instructor of Peiyang (NO.11303267).


\begin{thebibliography}{61}
\expandafter\if
x\csname natexlab\endcsname\relax\def\natexlab#1{#1}\fi
\expandafter\ifx\csname bibnamefont\endcsname\relax
  \def\bibnamefont#1{#1}\fi
\expandafter\ifx\csname bibfnamefont\endcsname\relax
  \def\bibfnamefont#1{#1}\fi
\expandafter\ifx\csname citenamefont\endcsname\relax
  \def\citenamefont#1{#1}\fi
\expandafter\ifx\csname url\endcsname\relax
  \def\url#1{\texttt{#1}}\fi
\expandafter\ifx\csname urlprefix\endcsname\relax\def\urlprefix{URL }\fi
\providecommand{\bibinfo}[2]{#2}
\providecommand{\eprint}[2][]{\url{#2}}

\bibitem[{\citenamefont{{H. Y. Yu}}(2012)}]{wxj1}
\bibinfo{author}{\bibnamefont{{Wang. Q. H, Zadeh. K. K,	Kis. A,	 Coleman. J. N, and Strano. M. S}}}, \bibinfo{year}{2012} \bibinfo{journal}{Nature Nanotech.} \textbf{\bibinfo{volume}{7}} \bibinfo{pages}{699}

\bibitem[{\citenamefont{{H. Y. Yu}}(2015)}]{wx1}
\bibinfo{author}{\bibnamefont{{Yu. H. Y, Cui. X. D, Xu. X. D, and Yao. W}}}, \bibinfo{year}{2015} \bibinfo{journal}{Natl. Sci. Rev.} \textbf{\bibinfo{volume}{2}} \bibinfo{pages}{57}


\bibitem[{\citenamefont{{A. V. Kolobov}}(2016)}]{wx2}
\bibinfo{author}{\bibnamefont{{Kolobov. A. V and Tominaga. J}}}, \bibinfo{journal} \bibinfo{year}{2016} {Two-Dimensional Transition-Metal Dichalcogenides, Springer Series in Materials Science 239} \bibinfo{pages}{321-361}
.

\bibitem[{\citenamefont{{T. Cheiwchanchamnangij}}(2012)}]{wx3}
\bibinfo{author}{\bibnamefont{{Cheiwchanchamnangij. T, and Lambrecht. W. R. L}}}, \bibinfo{year}{2012} \bibinfo{journal}{Phys. Rev. B} \textbf{\bibinfo{volume}{85}} \bibinfo{pages}{205302}


\bibitem[{\citenamefont{{T. C. Berkelbach}}(2013)}]{wx4}
\bibinfo{author}{\bibnamefont{{Berkelbach. T. C, Hybertsen. M. S, and Reichman. D. R}}}, \bibinfo{year}{2013} \bibinfo{journal}{Phys. Rev. B} \textbf{\bibinfo{volume}{88}} \bibinfo{pages}{045318}


\bibitem[{\citenamefont{{D. Y. Qiu}}(2013)}]{wx5}
\bibinfo{author}{\bibnamefont{{Qiu. D. Y, da Jornada. F. H, and Louie. S. G}}}, \bibinfo{year}{2013} \bibinfo{journal}{Phys. Rev. Lett.} \textbf{\bibinfo{volume}{111}} \bibinfo{pages}{216805}


\bibitem[{\citenamefont{{M. M. Ugeda}}(2014)}]{wx6}
\bibinfo{author}{\bibnamefont{{Ugeda. M. M, Bradley. A. J, Shi. S. F , da Jornada. F. H, Zhang. Y, Qiu. D. Y, Ruan. W, Mo. S. K, Hussain. Z, Shen. Z. X, Wang. F, Louie. S. G, and Crommie. M. F}}}, \bibinfo{year}{2014} \bibinfo{journal}{Nat. Mater.} \textbf{\bibinfo{volume}{13}} \bibinfo{pages}{1091}


\bibitem[{\citenamefont{{J. H. Choi}}(2015)}]{wx7}
\bibinfo{author}{\bibnamefont{{ Choi. J. H, Cui. P, Lan. H. P, and Zhang. Z. Y}}}, \bibinfo{year}{2015} \bibinfo{journal}{Phys. Rev. Lett.} \textbf{\bibinfo{volume}{115}} \bibinfo{pages}{066403}


\bibitem[{\citenamefont{{H. M. Hill}}(2015)}]{wx8}
\bibinfo{author}{\bibnamefont{{Hill. H. M, Rigosi. A. F, Roquelet. C , Chernikov. A, Berkelbach. T. C, Reichman. D. R, Hybertsen. M. S, Brus. L. E, and Heinz. T. F}}}, \bibinfo{year}{2015} \bibinfo{journal}{Nano. Lett.} \textbf{\bibinfo{volume}{15}} \bibinfo{pages}{2992}

\bibitem[{\citenamefont{{H. Y. Shi}}(2013)}]{wx9}
\bibinfo{author}{\bibnamefont{{Shi. H. Y, Yan. R. S, Bertolazzi. S, Brivio. J, Gao. B, Kis. A, Jena. D, Xing. H. L. G, and Huang. L. B }}}, \bibinfo{year}{2013} \bibinfo{journal}{ACS Nano} \textbf{\bibinfo{volume}{7}} \bibinfo{pages}{1072}


\bibitem[{\citenamefont{{J. W. Kim}}(2014)}]{wx10}
\bibinfo{author}{\bibnamefont{{Kim. J. W, Hong. X. P, Jin. C. H, Shi. S. F, Chang. C. Y. S, Chiu. M. H, Li. L. J, Wang. F}}}, \bibinfo{year}{2014} \bibinfo{journal}{Science} \textbf{\bibinfo{volume}{346}} \bibinfo{pages}{1205}


\bibitem[{\citenamefont{{A. T. Hanbicki}}(2015)}]{wx11}
\bibinfo{author}{\bibnamefont{{Hanbicki. A. T, Currie. M, Kioseoglou. G, Friedman. A. L, and Jonker. B. T}}}, \bibinfo{year}{2015} \bibinfo{journal}{Solid. State. Commun.} \textbf{\bibinfo{volume}{203}} \bibinfo{pages}{16}


\bibitem[{\citenamefont{{X. X. Zhang}}(2015)}]{wx12}
\bibinfo{author}{\bibnamefont{{Zhang. X. X, You. Y. M, Zhao. S. Y. F , and Heinz. T. F}}}, \bibinfo{year}{2015} \bibinfo{journal}{Phys. Rev. Lett.} \textbf{\bibinfo{volume}{115}} \bibinfo{pages}{257403}


\bibitem[{\citenamefont{{K. L. He}}(2014)}]{wx13}
\bibinfo{author}{\bibnamefont{{He. K. L, Kumar. N, Zhao. L, Wang. Z. F, Mak. K. F, Zhao. H, and Shan. J}}}, \bibinfo{year}{2014}
\bibinfo{journal}{Phys. Rev. Lett.} \textbf{\bibinfo{volume}{113}} \bibinfo{pages}{026803}

\bibitem[{\citenamefont{{J. Li}}(2015)}]{wx14}
\bibinfo{author}{\bibnamefont{{Li. J, Zhong. Y. L, and Zhang. D}}}, \bibinfo{year}{2015} \bibinfo{journal}{J. Phys.: Condens. Matter} \textbf{\bibinfo{volume}{27}} \bibinfo{pages}{315301}

\bibitem[{\citenamefont{{B. R. Zhu}}(2015)}]{wx15}
\bibinfo{author}{\bibnamefont{{Zhu. B. R, Chen. X, and Cui. X. D}}}, \bibinfo{year}{2015} \bibinfo{journal}{Sci. Rep.} \textbf{\bibinfo{volume}{5}} \bibinfo{pages}{9218}


\bibitem[{\citenamefont{{T. C. Berkelbach}}(2013)}]{wx16}
\bibinfo{author}{\bibnamefont{{Berkelbach. T. C, Hybertsen. M. S, and Reichman. D. R,}}}, \bibinfo{year}{2013} \bibinfo{journal}{Phys. Rev. B} \textbf{\bibinfo{volume}{88}} \bibinfo{pages}{045318}


\bibitem[{\citenamefont{{A. Chernikov}}(2014)}]{wx17}
\bibinfo{author}{\bibnamefont{{Chernikov. A, Berkelbach. T. C, Hill. H. M, Rigosi. A, Li. Y. L, Aslan. O. B, Reichman. D. R, Hybertsen. M. S, and Heinz. T. F}}}, \bibinfo{year}{2014} \bibinfo{journal}{Phys. Rev. Lett.} \textbf{\bibinfo{volume}{113}} \bibinfo{pages}{076802}


\bibitem[{\citenamefont{{T. Olsen}}(2016)}]{wx18}
\bibinfo{author}{\bibnamefont{{Olsen. T, Latini. S, Rasmussen. F, and Thygesen. K. S}}}, \bibinfo{year}{2016} \bibinfo{journal}{Phys. Rev. Lett.} \textbf{\bibinfo{volume}{116}} \bibinfo{pages}{056401}


\bibitem[{\citenamefont{{M. Trushin}}(2016)}]{wx19}
\bibinfo{author}{\bibnamefont{{Trushin. M, Goerbig. M. O, and Belzig. W}}}, \bibinfo{year}{2016} \bibinfo{journal}{Phys. Rev. B} \textbf{\bibinfo{volume}{94}} \bibinfo{pages}{041301}


\bibitem[{\citenamefont{{A. Thilagam}}(2014)}]{wx20}
\bibinfo{author}{\bibnamefont{{Thilagam. A}}}, \bibinfo{year}{2014} \bibinfo{journal}{J. Appl. Phys.} \textbf{\bibinfo{volume}{323}} \bibinfo{pages}{053523}


\bibitem[{\citenamefont{{K. Kaasbjerg}}(2012)}]{wx21}
\bibinfo{author}{\bibnamefont{{Kaasbjerg. K, Thygesen. K. S, and Jacobsen. K. W}}}, \bibinfo{year}{2012} \bibinfo{journal}{Phys. Rev. B} \textbf{\bibinfo{volume}{85}} \bibinfo{pages}{115317}


\bibitem[{\citenamefont{{K. Kaasbjerg}}(2014)}]{wx22}
\bibinfo{author}{\bibnamefont{{Kaasbjerg. K, Thygesen. K. S, and Jacobsen. K. W}}}, \bibinfo{year}{2014} \bibinfo{journal}{Phys. Rev. B} \textbf{\bibinfo{volume}{90}} \bibinfo{pages}{165436}


\bibitem[{\citenamefont{{A. Thilagam}}(2016)}]{wx23}
\bibinfo{author}{\bibnamefont{{Thilagam. A}}}, \bibinfo{year}{2016} \bibinfo{journal}{J. Appl. Phys.} \textbf{\bibinfo{volume}{120}} \bibinfo{pages}{124306}


\bibitem[{\citenamefont{{I. T. Lin}}(2013)}]{wx24}
\bibinfo{author}{\bibnamefont{{Lin. I. T, and Liu. J. M}}}, \bibinfo{year}{2013} \bibinfo{journal}{Appl. Phys. Lett.} \textbf{\bibinfo{volume}{103}} \bibinfo{pages}{081606}

\bibitem[{\citenamefont{{R. Rengel}}(2014)}]{wx25}
\bibinfo{author}{\bibnamefont{{Rengel. R, Pascual. E, and Mart\'{i}n. M. J}}}, \bibinfo{year}{2014} \bibinfo{journal}{Appl. Phys. Lett.} \textbf{\bibinfo{volume}{104}} \bibinfo{pages}{233107}


\bibitem[{\citenamefont{{A. Mogulkoc}}(2016)}]{wx26}
\bibinfo{author}{\bibnamefont{{Mogulkoc. A, Mogulkoc. Y, Rudenko. A. N, and Katsnelson. M. I}}}, (\bibinfo{year}{2016}) \bibinfo{journal}{Phys. Rev. B} \textbf{\bibinfo{volume}{93}}  \bibinfo{pages}{085417}

\bibitem[{\citenamefont{{T. D. Lee}}(1953)}]{wx27}
\bibinfo{author}{\bibnamefont{{Lee. T. D, Low. F. E, and Pines. D }}}, \bibinfo{year}{1953} \bibinfo{journal}{Phys. Rev.} \textbf{\bibinfo{volume}{90}} \bibinfo{pages}{297}


\bibitem[{\citenamefont{{J. Pollmann}}(1977)}]{wx28}
\bibinfo{author}{\bibnamefont{{Pollmann. J and B$\ddot{u}$ttner. H }}}, \bibinfo{year}{1977} \bibinfo{journal}{Phys. Rev. B} \textbf{\bibinfo{volume}{16}} \bibinfo{pages}{4480}


\bibitem[{\citenamefont{{A. Ramasubramaniam}}(2012)}]{wx29}
\bibinfo{author}{\bibnamefont{{Ramasubramaniam. A}}}, \bibinfo{year}{2012} \bibinfo{journal}{Phys. Rev. B} \textbf{\bibinfo{volume}{86}} \bibinfo{pages}{115409}


\bibitem[{\citenamefont{{A. Thilagam}}(2015)}]{wx30}
\bibinfo{author}{\bibnamefont{{Thilagam. A}}}, \bibinfo{year}{2015} \bibinfo{journal}{Physica B} \textbf{\bibinfo{volume}{464}} \bibinfo{pages}{44}


\bibitem[{\citenamefont{{A. Konar}}(2014)}]{wx31}
\bibinfo{author}{\bibnamefont{{Jin. Z. H, Li. X. D, Mullen. J. T, and Kim. K. W}}}, \bibinfo{year}{2014} \bibinfo{journal}{Phys. Rev. B} \textbf{\bibinfo{volume}{90}} \bibinfo{pages}{045422}


\bibitem[{\citenamefont{{T. Sohier}}(2016)}]{wx32}
\bibinfo{author}{\bibnamefont{{Sohier. T, Calandra. M, and Mauri. F }}}, \bibinfo{year}{2016} \bibinfo{journal}{Phys. Rev. B}
 \textbf{\bibinfo{volume}{94}} \bibinfo{pages}{085415}
 

\bibitem[{\citenamefont{{C. H. Jin}}(2016)}]{wx33}
\bibinfo{author}{\bibnamefont{{Jin. C. H, Kim. J, Suh. J, Shi. Z. W, Chen. B, Fan. X, Kam. M, Watanabe. K, Taniguchi. T, Tongay. S, Zettl. A, Wu. J. Q, and Wang. F}}}, \bibinfo{year}{2016} \bibinfo{journal}{Nat. Phys.} \textbf{\bibinfo{volume}{10}}, \bibinfo{pages}{1038}
 

\end{thebibliography}
\end{document}